\documentclass[prb,showpacs,floatfix,twocolumn,byrevtex,superscriptaddress]{revtex4-1}
\usepackage[english]{babel}
\usepackage{amsmath}
\usepackage{graphicx}
\usepackage{float}
\usepackage{bm}
\usepackage{multirow}
\usepackage{dcolumn}

\newcommand{\icm}{\ensuremath{~\textrm{cm}^{-1}}}
\newcommand{\BSLCO}{Bi$_2$Sr$_{2-x}$La$_x$CuO$_6$}

\newcolumntype{.}{D{.}{.}{-1}}

\begin{document}

\bibliographystyle{apsrev}

\title{Doping evolution of the optical scattering rate and effective mass of \BSLCO}

\author{Y. M. Dai}
\affiliation{Beijing National Laboratory for Condensed Matter Physics, National Laboratory for Superconductivity, Institute of Physics, Chinese Academy of Sciences, P.O. Box 603, Beijing 100190, China}
\affiliation{LPEM, ESPCI-ParisTech, CNRS, UPMC, 10 rue Vauquelin, F-75231 Paris Cedex 5, France}

\author{B. Xu}
\affiliation{Beijing National Laboratory for Condensed Matter Physics, National Laboratory for Superconductivity, Institute of Physics, Chinese Academy of Sciences, P.O. Box 603, Beijing 100190, China}

\author{P. Cheng}
\affiliation{Beijing National Laboratory for Condensed Matter Physics, National Laboratory for Superconductivity, Institute of Physics, Chinese Academy of Sciences, P.O. Box 603, Beijing 100190, China}

\author{H. Q. Luo}
\affiliation{Beijing National Laboratory for Condensed Matter Physics, National Laboratory for Superconductivity, Institute of Physics, Chinese Academy of Sciences, P.O. Box 603, Beijing 100190, China}

\author{H. H. Wen}
\affiliation{Beijing National Laboratory for Condensed Matter Physics, National Laboratory for Superconductivity, Institute of Physics, Chinese Academy of Sciences, P.O. Box 603, Beijing 100190, China}
\affiliation{National Laboratory of Solid State Microstructures and Department of Physics, Nanjing University, Nanjing 210093, China}

\author{X. G. Qiu}
\affiliation{Beijing National Laboratory for Condensed Matter Physics, National Laboratory for Superconductivity, Institute of Physics, Chinese Academy of Sciences, P.O. Box 603, Beijing 100190, China}

\author{R. P. S. M. Lobo}
\email[]{lobo@espci.fr}
\affiliation{LPEM, ESPCI-ParisTech, CNRS, UPMC, 10 rue Vauquelin, F-75231 Paris Cedex 5, France}

\date{\today}

\begin{abstract}
We determined the optical conductivity of \BSLCO\ at dopings covering the phase diagram from the underdoped to the overdoped regimes. The frequency dependent scattering rate shows a pseudogap extending into the overdoped regime. We found that the effective mass enhancement calculated from the optical conductivity is constant throughout the phase diagram. Conversely, the effective optical charge density varies almost linearly with doping. Our results suggest that the low frequency electrodynamics of \BSLCO\ is not strongly affected by the long range Mott transition.
\end{abstract}

\maketitle

%
%

Hole doped cuprates are strongly correlated systems where the Coulomb repulsion prevents electrons from hoping between nearest Cu atoms yielding a zero doping ground state that is a Mott-Hubbard insulator instead of a half-filled metal. Nevertheless, increasing the hole concentration $p$ replaces this Mott insulator by exotic normal state metallic phases. For a range of dopings, a superconducting dome appears. The understanding of the superconducting mechanism passes through the comprehension of the normal state phases and, in particular, how the physical properties of cuprate superconductors evolve with the strength of correlations.

\citet{Millis2005} showed that comparing the optical conductivity to LDA band structure calculations is a valuable tool to determine the degree of correlation in cuprates. In particular they showed that both electron and hole doped cuprates have a high degree of correlations that decreases by a factor of 3 almost linearly with doping. Despite this large change in the correlation strength, \citet{Padilla2005} showed that in YBa$_2$Cu$_3$O$_{6+\delta}$ and La$_{2-x}$Sr$_x$CuO$_4$ the effective mass enhancement ($m/m^\star$) is the same over a large portion of the phase diagram. In their analysis, they estimate $m/m^\star$ from the weight of the coherent Drude peak below $\sim 100$~meV, although they argue that the result should be qualitatively the same if one considers the whole conduction band spectral weight. Hence the Mott transition and transformations among the metallic phases would not be driven by a mass divergence but by charge density modifications alone. This picture is compatible with photoemission finding a charge density varying linearly with doping\cite{Yoshida2003,Hashimoto2008,Pan2009} but a nodal Fermi velocity independent of doping.\cite{Zhou2003,Kondo2006} Photoemission also finds that at low dopings the Fermi surface is confined to the nodal regions and it evolves into a full arc close to optimal doping.\cite{Hashimoto2008}

Single layer cuprate superconductors derived from Bi$_2$Sr$_2$CuO$_6$ (Bi2201) are a model system for the study of normal state properties throughout the phase diagram.\cite{Eisaki2004} Their relatively low $T_c$ allows for measurements at low temperatures in the normal state. Its lone CuO$_2$ plane avoids effects of interplane coupling. Substitution of Sr by La (\BSLCO) yields the superconductor with the lowest influence of disorder and hence the highest $T_c^{Max} \sim 33$~K in the Bi2201 family.\cite{Fujita2005} It spans hole concentrations from extreme (non superconducting) underdoped to strongly overdoped samples allowing to probe a large range of exotic phases such as unconventional magnetic order, a pseudogap state, and high-$T_c$ superconductivity. 

Little optical data exists on single layer Bi2201 based superconductors. \citet{Romero1992} early infrared data showed that the scattering rate obtained from optics is compatible with dc transport data and depended linearly on temperature. \citet{Tsvetkov1997} found a pseudogap in the normal state of oxygen doped Bi2201 ($T_c = 20$~K) and concluded for a system with two coupled conducting channels. \citet{Lupi2000} measurements on a superconducting oxygen doped Bi2201 film ($T_c = 20$~K) indicated a multicomponent optical conductivity with the coexistence of free and bound carriers, making it difficult to observe a pseudogap spectral weight signature. This latter group also obtained more recent data on a few different La concentration levels in the underdoped regime of Bi2201.\cite{Lupi2009} They find the development of a Drude-like peak with doping compatible with photoemission data, stressing the important contribution of the nodal quasiparticles in reciprocal space to the optical conductivity.

In this note, we measured the optical conductivity of \BSLCO\ at several dopings from the underdoped to the overdoped regimes. Utilizing the full conduction band spectral weight and in agreement with \citet{Padilla2005}, we find a constant optical effective mass enhancement whereas the charge density changes by a factor of 10. We find a pseudogap extending into the overdoped regime, but with no obvious signature in the mass enhancement. Our results support the view that the dynamical response of quasiparticles in cuprates is local and is not strongly influenced by the Mott transition.

%
%
%
\begin{table}
\begin{center}
\caption{Nominal La concentration, hole doping concentration determined from \citet{Ando2000}, critical temperatures and plasma frequencies using Eq.~\ref{EqSumRule} for our \BSLCO\ samples.}
\begin{ruledtabular}
\begin{tabular}{....}
\multicolumn{1}{c}{$x$} & \multicolumn{1}{c}{$p$} & \multicolumn{1}{c}{$T_c$ (K)} & \multicolumn{1}{c}{$\Omega_p$ (cm$^{-1}$)}\\
\hline
0.1 & 0.198	& 21 & 14\,000 \\
0.2 & 0.184	& 28 & 13\,000 \\
0.4 & 0.16 & 33 & 12\,100	\\
0.6 & 0.129 &	25 & 12\,000 \\
0.7 & 0.115	& 16 & 10\,500 \\
0.8 & 0.109	& 11 & 10\,800 \\
1.0 & 0.03	&  0 & 2\,900 \\
\end{tabular}
\end{ruledtabular}
\label{Tab1}
\end{center}
\end{table}
High quality single crystals of \BSLCO\ were grown by optical traveling-solvent floating-zone with nominal La concentrations ranging from 0.1 to 1.0 (Ref. \onlinecite{Luo2008}). Our crystals have sharp magnetic and resistive transitions. We obtained the hole concentration ($p$) from $T_c$ utilizing the relation $T_c / T_c^{Max} = 1 - 255 (p - 0.16)^2$ determined specifically for \BSLCO\ by \citet{Ando2000} Our range of dopings cover the phase diagram from the highly overdoped ($x=0.1$; $p\sim0.2$; $T_c = 21$~K) to the underdoped ($x=1.0$; $p\sim0.03$; $T_c = 0$) regimes. The maximum $T_c = 33$~K was measured in the $x=0.4$ samples and we defined this as the optimal doping $p=0.16$. Table~\ref{Tab1} summaryzes our samples characteristics.

Near normal incidence reflectivity spectra from 20 to 20\,000\icm\ were measured at several temperatures on cleaved surfaces of crystals with typically $2 \times 2 \textrm{ mm}^2$ utilizing Bruker IFS113 and IFS66v spectrometers. The absolute reflectivity, with an accuracy better than 0.5\%, was obtained with an \emph{in situ} gold overfilling technique.\cite{Homes1993} We concentrate our analysis at data obtained at 100~K. This temperature is low enough to produce a well established Drude-like peak and access the pseudo-gap and the metallic normal state phases. Yet, this temperature is high enough to prevent eventual effects of superconducting fluctuations in the optical conductivity. 

\begin{figure}[htb]
  \includegraphics[width=7.5cm]{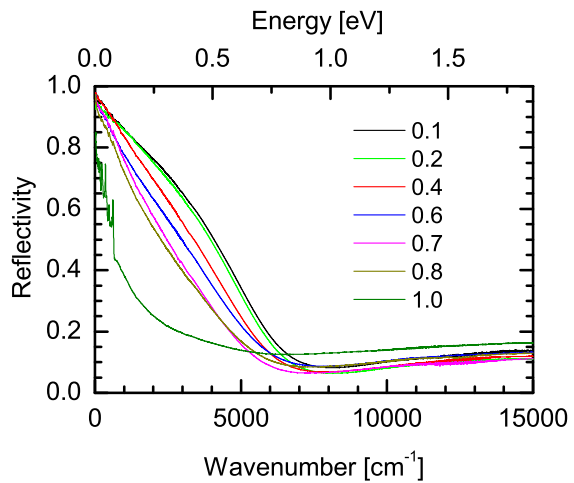}
  \caption{(color online) Reflectivity of \BSLCO\ measured at 100 K spanning the phase diagram from the underdoped regime to the overdoped region. 
}
  \label{Fig1}
\end{figure}
Figure~\ref{Fig1} shows the measured 100 K reflectivity spectra for all dopings. In the most underdoped sample ($x = 1.0$) we observe phonon peaks below 600\icm\ dominating the reflectivity. Nevertheless, a negative slope, characteristic of an overdamped Drude-like response, is present attesting the conductive character of this sample. Upon increasing hole doping, phonon peaks are quickly screened by a stronger response of mobile carriers and a clearer Drude-like term appears. Although the simple Drude response cannot account for the optical response of cuprates, we can approximate a plasma edge to the minimum in the reflectivity, around 8\,000\icm\ (1 eV). We can see that this minimum also increases with doping, indicating an increase in the charge density.

%
%
To quantify our data, we calculated the optical conductivity $[\sigma(\omega)]$ from the reflectivity through Kramers-Kronig analysis. At low frequencies we utilized a Hagen-Rubens ($1 - A \sqrt{\omega}$) extrapolation whereas at high frequencies we took a constant reflectivity to 300\,000\icm\ (37.5 eV) followed by a $\omega^{-4}$ free electron termination. 

\begin{figure}[htb]
  \includegraphics[width=7.5cm]{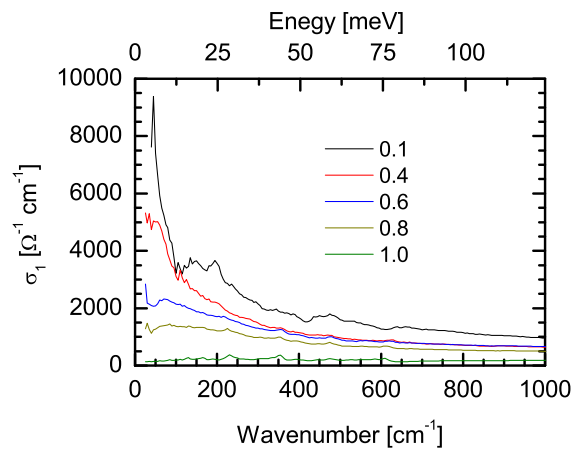}
  \caption{(color online) Far-infrared real part of the optical conductivity for \BSLCO\ at 100 K from the underdoped ($x=1.0$) to the overdoped regime ($x=0.1$). A Drude-like peak grows monotonously with increasing doping. 
}
  \label{Fig2}
\end{figure}
Figure \ref{Fig2} shows the real part of the optical conductivity $[\sigma_1(\omega)]$ for our samples. For clarity, samples with $x=0.2$ and $x=0.7$, which have responses very close to $x=0.1$ and $x=0.8$, have been omitted. Except for phonons, our most underdoped sample ($x = 1.0$) has an almost flat, incoherent, frequency independent far-infrared $\sigma_1$. With increasing doping, a Drude-like peak develops increasing the low frequency spectral weight, extending the behavior observed by \citet{Lupi2009} into the overdoped regime. 

We analyzed the optical conductivity with the extended Drude model proposed by \citet{Allen1971}. Here the scattering rate ($1/\tau$) in the Drude model is allowed to have a frequency dependence. Causality relations then imply that a mass renormalization term ($1+\lambda = m^\star / m$) must also depend on the frequency. Defining the plasma frequency as $\Omega_p$ and taking the vacuum impedance as $Z_0 = 377~\Omega$, the extended optical conductivity becomes:
\begin{equation}
\sigma(\omega) = \frac{2\pi}{Z_0}\frac{\Omega_p^2}{\tau^{-1}(\omega) - i \omega\left[1+\lambda(\omega)\right]}.
\label{EqExtDrude}
\end{equation}

Inverting Eq.~\ref{EqExtDrude} allows one to calculate the experimental curves for $1/\tau$ and $1+\lambda$ as:
\begin{align}
\frac{1}{\tau(\omega)} &= \frac{2\pi}{Z_0}\Omega_p^2\textrm{ Re}\left(\frac{1}{\sigma(\omega)}\right); \label{EqTau}\\
1 + \lambda(\omega) &= \frac{1}{\omega}\frac{2\pi}{Z_0}\Omega_p^2\textrm{Im}\left(\frac{1}{\sigma(\omega)}\right). \label{EqLambda}
\end{align}
The tricky point in Eqs.~\ref{EqTau} and \ref{EqLambda} is the determination of the value of the plasma frequency. Here, we utilized the optical conductivity partial sum-rule that states:
\begin{equation}
\int_0^{\omega_c} \sigma_1(\omega) d\omega = \frac{\pi^2}{Z_0} \Omega_p^2,
\label{EqSumRule}
\end{equation}
where $\omega_c$ is a cut-off frequency typical of the conduction band width. We took $\omega_c =10\,000\icm$ (1.25 eV) which corresponds to a minimum between the Drude-like peak, characteristic of the conduction band, and interband transitions that start at 2 eV (Ref.~\onlinecite{Terasaki1990}). This minimum is not well defined for the most underdoped sample. In that case, integration was done after fitting $\sigma_1$ to a Drude-Lorentz dielectric function and subtracting phonons and mid-infrared transitions. The uncertainty on which mid-infrared oscillators should one eliminate in the $x = 1.0$ sample produces a larger uncertainty in $\Omega_p$ and on the shape and magnitude of the optical scattering rate and mass enhancement. Our values for $\Omega_p$ are shown in Tab.~\ref{Tab1}.

%
%

%
\begin{figure}[htb]
  \includegraphics[width=7.5cm]{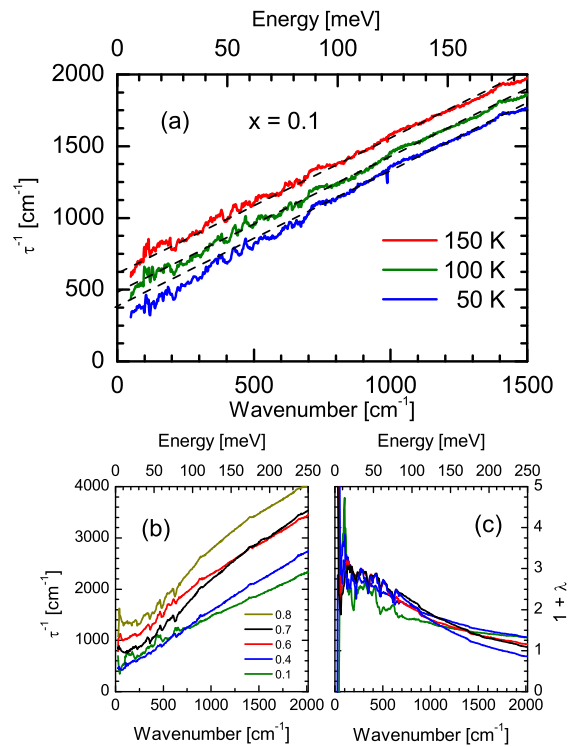}
  \caption{(color online) (a) Frequency dependent scattering rate at 150, 100 and 50 K in the most overdoped sample. For clarity, phonons have been subtracted. All dashed straight lines have the same slope. At 50 K, the linear behavior of $\tau^{-1}$ breaks down below $\sim 700 \icm$, indicative of a pseudogap opening. (b) Scattering rate at 100 K for several doping levels. Once again, the pseudogap manifests itself as a departure from linearity in the low frequency behavior of $\tau^{-1}$ observed in the underdoped samples (higher values of $x$). (c) Mass enhancement at 100 K for the same samples shown in (b).
}
  \label{Fig3}
\end{figure}
The outcome of Eqs.~\ref{EqTau} and \ref{EqLambda} on a relatively large frequency range is shown in Fig.~\ref{Fig3}. Panel (a) shows $\tau^{-1}(\omega)$ at 150, 100 and 50 K for our most overdoped ($x = 0.1$) sample. For clarity, small phonon peaks have been subtracted. One can see that at 150 and 100 K, as indicated by the dashed lines, $\tau^{-1}$ is linear in frequency in the whole range shown. At 50 K, $\tau^{-1}$ follows the same linear trend as the one observed at higher temperatures but only above $\sim 700 \icm$. The departure from linearity below that frequency is the hallmark of the pseudogap opening.\cite{Puchkov1996} This sample is the one showing the smallest departure from linearity in $\tau^{-1}$. Repeating this analysis for all samples, we could estimate the pseudogap opening temperature.

Figure~\ref{Fig3} (b) shows $\tau^{-1}(\omega)$ for several dopings at 100~K. The scattering rate at this temperature is linear up to about 2\,000\icm\ (250 meV) for samples with $x = 0.1$ and 0.2 (not shown for clarity). For higher La contents, marching towards the underdoped regime, $1 / \tau$ is supressed at low frequencies, departing from the linear behavior. Panel (c) depicts the mass enhancement for the same samples shown in (b). All curves are very similar to each other, despite the use of plasma frequency values that vary by 40~\% (see Tab.~\ref{Tab1}) and, hence, charge densities that change by a factor of 2 for the shown dopings. The data only become significantly different in the $\sim 4000\icm$ (0.5 eV) region where the mass enhancement becomes negative due to effects of a lower plasma edge and the tail of interband transitions. Because of the aforementioned larger error for the $x = 0.1$, we are not showing these optical functions for that sample. Nevertheless, its very low frequency response produces useful values, as we discuss in the following.

\begin{figure}[htb]
  \includegraphics[width=7.5cm]{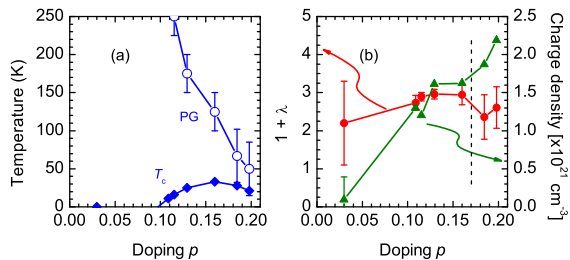}
  \caption{(color online) (a) Phase diagram of \BSLCO\ showing the pseudogap (extracted from $1/\tau$) as open circles and the critical temperatures (from the resistive transition) as solid diamonds. The pseudogap temperature for the two most underdoped samples are above room temperature and are not shown. (b) Doping evolution, at 100 K, of the low frequency mass enhancement (solid circles, left hand side scale) and the effective charge density (solid triangles, right hand side scale) obtained by integrating $\sigma_1$ up to 10000\icm (1.25 eV). At 100 K, the pseudogap is present at all dopings up to the vertical dashed line. Unless indicated otherwise, error bars are smaller than symbol sizes.
}
  \label{Fig4}
\end{figure}
Figure \ref{Fig4} summarizes the main properties of \BSLCO\ as a function of doping. In panel (a), the dome formed by the solid diamonds (labeled $T_c$) shows the superconducting transition temperatures obtained from the dc resistivity of our samples. The monotonically decreasing curve (open circles, labeled PG) defines the pseudogap line. The two most underdoped samples ($x = 0.8$ and 1.0) have pseudogap temperatures above 300~K that cannot be determined from our data. The error bars come from our temperature step (25 K) and deciding exactly which temperature shows $\tau^{-1}$ departing from the linear behavior. It is remarkable that a pseudogap is present in all of our dopings and not just in the underdoped region, in agreement with NMR, tunneling and photoemission data.\cite{Yurgens2003,Zheng2005,Ma2008,Kawasaki2010}

The solid circles in Fig.~\ref{Fig4} (b), reporting to the left hand side ordinate, show the low frequency effective mass enhancement calculated from the 100 K data in Fig.~\ref{Fig3}. Values for $1 + \lambda$ are obtained by averaging the data in the 100--300\icm\ range and taking the standard deviation as the error bar. This operation is more complex in the extremely underdoped sample as the electronic contribution is small compared to phonon and mid-infrared bands. Because of the uncertainties mentioned above in the determination of the mobile carrier contribution to $\sigma_1$ in the $x = 1.0$ sample a larger error bar exists for its $1 + \lambda$ value. Nevertheless, within error bars, the effective mass is constant over the phase diagram. More important, even considering the larger error bar for $x = 1.0$, no sign of divergence of $1 + \lambda$ at the insulating phase is present. 

Figure \ref{Fig4} (b) also shows the effective charge density ($n$) calculated at 100 K from the plasma frequency using the relation $\Omega_p^2 = n e^2 / \varepsilon_0 m$, where $e$ and $m$ are the carrier charge and mass, respectively and $\varepsilon_0$ is the vacuum permittivity. We utilized the bare electronic mass for $m$ but, as $1 + \lambda$ is constant over the phase diagram, the result of utilizing the effective mass is simply to produce a charge density about 3 times larger. The relative values of $n$, however, remain unchanged. The optically determined $n$ increases almost linearly with doping, by a factor of 10 when considering the most underdoped sample, reproducing a similar trend from photoemission.\cite{Hashimoto2008} At 100 K, the pseudogap is present at all dopings up to the vertical dashed line. The opening of the pseudo gap has no remarkable effect in $1 + \lambda$ or $n$. 

Comparing $1 + \lambda$ with $n$ shows that \BSLCO\ behaves similarly to YBa$_2$Cu$_3$O$_{6+\delta}$ and La$_{2-x}$Sr$_x$CuO$_4$ in the sense that the metal-insulator transition seems to be induced by a charge density decrease alone, not by an effective mass divergence. The dynamical response of \BSLCO, portrayed by its low frequency mass enhancement is local and not influenced by the long range Mott insulator transition. We may also speculate that since $1 + \lambda$ is constant over the phase diagram and $n$ varies almost linearly with doping, $T_c$ seems to be strongly associated with the carrier density and weakly depends on the electron-boson coupling strength.
 
%
%

In summary, we showed the optical conductivity of \BSLCO\ at several dopings spanning a large portion of the phase diagram from the underdoped to the overdoped regimes. An extended Drude analysis of the optical conductivity indicates a pseudogap present in all dopings measured, in particular deep into the overdoped regime. The optical effective mass enhancement is constant throughout the phase diagram in contrast to a charge density that changes by a factor of 10. These observations indicate that the dynamical response of quasiparticles in \BSLCO\ is local and are not strongly influenced by the Mott transition.

%
%

We would like to acknowledge the financial support from the Science and Technology Service of the French Embassy in China. Work in Beijing was supported by the MOST and the National Science Foundation of China. Work in Paris was supported by the ANR under Grant No. BLAN07-1-183876 GAPSUPRA.
%
%
\bibliography{Dai_Bi2201}

\begin{thebibliography}{23}
\expandafter\ifx\csname natexlab\endcsname\relax\def\natexlab#1{#1}\fi
\expandafter\ifx\csname bibnamefont\endcsname\relax
  \def\bibnamefont#1{#1}\fi
\expandafter\ifx\csname bibfnamefont\endcsname\relax
  \def\bibfnamefont#1{#1}\fi
\expandafter\ifx\csname citenamefont\endcsname\relax
  \def\citenamefont#1{#1}\fi
\expandafter\ifx\csname url\endcsname\relax
  \def\url#1{\texttt{#1}}\fi
\expandafter\ifx\csname urlprefix\endcsname\relax\def\urlprefix{URL }\fi
\providecommand{\bibinfo}[2]{#2}
\providecommand{\eprint}[2][]{\url{#2}}

\bibitem[{\citenamefont{Millis et~al.}(2005)\citenamefont{Millis, Zimmers,
  Lobo, Bontemps, and Homes}}]{Millis2005}
\bibinfo{author}{\bibfnamefont{A.~J.} \bibnamefont{Millis}},
  \bibinfo{author}{\bibfnamefont{A.}~\bibnamefont{Zimmers}},
  \bibinfo{author}{\bibfnamefont{R.~P. S.~M.} \bibnamefont{Lobo}},
  \bibinfo{author}{\bibfnamefont{N.}~\bibnamefont{Bontemps}}, \bibnamefont{and}
  \bibinfo{author}{\bibfnamefont{C.~C.} \bibnamefont{Homes}},
  \bibinfo{journal}{Phys. Rev. B} \textbf{\bibinfo{volume}{72}},
  \bibinfo{pages}{224517} (\bibinfo{year}{2005}).

\bibitem[{\citenamefont{Padilla et~al.}(2005)\citenamefont{Padilla, Lee, Dumm,
  Blumberg, Ono, Segawa, Komiya, Ando, and Basov}}]{Padilla2005}
\bibinfo{author}{\bibfnamefont{W.~J.} \bibnamefont{Padilla}},
  \bibinfo{author}{\bibfnamefont{Y.~S.} \bibnamefont{Lee}},
  \bibinfo{author}{\bibfnamefont{M.}~\bibnamefont{Dumm}},
  \bibinfo{author}{\bibfnamefont{G.}~\bibnamefont{Blumberg}},
  \bibinfo{author}{\bibfnamefont{S.}~\bibnamefont{Ono}},
  \bibinfo{author}{\bibfnamefont{K.}~\bibnamefont{Segawa}},
  \bibinfo{author}{\bibfnamefont{S.}~\bibnamefont{Komiya}},
  \bibinfo{author}{\bibfnamefont{Y.}~\bibnamefont{Ando}}, \bibnamefont{and}
  \bibinfo{author}{\bibfnamefont{D.~N.} \bibnamefont{Basov}},
  \bibinfo{journal}{Phys. Rev. B} \textbf{\bibinfo{volume}{72}},
  \bibinfo{pages}{060511} (\bibinfo{year}{2005}).

\bibitem[{\citenamefont{Yoshida et~al.}(2003)\citenamefont{Yoshida, Zhou,
  Sasagawa, Yang, Bogdanov, Lanzara, Hussain, Mizokawa, Fujimori, Eisaki
  et~al.}}]{Yoshida2003}
\bibinfo{author}{\bibfnamefont{T.}~\bibnamefont{Yoshida}},
  \bibinfo{author}{\bibfnamefont{X.~J.} \bibnamefont{Zhou}},
  \bibinfo{author}{\bibfnamefont{T.}~\bibnamefont{Sasagawa}},
  \bibinfo{author}{\bibfnamefont{W.~L.} \bibnamefont{Yang}},
  \bibinfo{author}{\bibfnamefont{P.~V.} \bibnamefont{Bogdanov}},
  \bibinfo{author}{\bibfnamefont{A.}~\bibnamefont{Lanzara}},
  \bibinfo{author}{\bibfnamefont{Z.}~\bibnamefont{Hussain}},
  \bibinfo{author}{\bibfnamefont{T.}~\bibnamefont{Mizokawa}},
  \bibinfo{author}{\bibfnamefont{A.}~\bibnamefont{Fujimori}},
  \bibinfo{author}{\bibfnamefont{H.}~\bibnamefont{Eisaki}},
  \bibnamefont{et~al.}, \bibinfo{journal}{Phys. Rev. Lett.}
  \textbf{\bibinfo{volume}{91}}, \bibinfo{pages}{027001}
  (\bibinfo{year}{2003}).

\bibitem[{\citenamefont{Hashimoto et~al.}(2008)\citenamefont{Hashimoto,
  Yoshida, Yagi, Takizawa, Fujimori, Kubota, Ono, Tanaka, Lu, Shen
  et~al.}}]{Hashimoto2008}
\bibinfo{author}{\bibfnamefont{M.}~\bibnamefont{Hashimoto}},
  \bibinfo{author}{\bibfnamefont{T.}~\bibnamefont{Yoshida}},
  \bibinfo{author}{\bibfnamefont{H.}~\bibnamefont{Yagi}},
  \bibinfo{author}{\bibfnamefont{M.}~\bibnamefont{Takizawa}},
  \bibinfo{author}{\bibfnamefont{A.}~\bibnamefont{Fujimori}},
  \bibinfo{author}{\bibfnamefont{M.}~\bibnamefont{Kubota}},
  \bibinfo{author}{\bibfnamefont{K.}~\bibnamefont{Ono}},
  \bibinfo{author}{\bibfnamefont{K.}~\bibnamefont{Tanaka}},
  \bibinfo{author}{\bibfnamefont{D.~H.} \bibnamefont{Lu}},
  \bibinfo{author}{\bibfnamefont{Z.-X.} \bibnamefont{Shen}},
  \bibnamefont{et~al.}, \bibinfo{journal}{Phys. Rev. B}
  \textbf{\bibinfo{volume}{77}}, \bibinfo{pages}{094516}
  (\bibinfo{year}{2008}).

\bibitem[{\citenamefont{Pan et~al.}(2009)\citenamefont{Pan, Richard, Xu,
  Neupane, Bishay, Fedorov, Luo, Fang, Wen, Wang et~al.}}]{Pan2009}
\bibinfo{author}{\bibfnamefont{Z.~H.} \bibnamefont{Pan}},
  \bibinfo{author}{\bibfnamefont{P.}~\bibnamefont{Richard}},
  \bibinfo{author}{\bibfnamefont{Y.~M.} \bibnamefont{Xu}},
  \bibinfo{author}{\bibfnamefont{M.}~\bibnamefont{Neupane}},
  \bibinfo{author}{\bibfnamefont{P.}~\bibnamefont{Bishay}},
  \bibinfo{author}{\bibfnamefont{A.~V.} \bibnamefont{Fedorov}},
  \bibinfo{author}{\bibfnamefont{H.}~\bibnamefont{Luo}},
  \bibinfo{author}{\bibfnamefont{L.}~\bibnamefont{Fang}},
  \bibinfo{author}{\bibfnamefont{H.~H.} \bibnamefont{Wen}},
  \bibinfo{author}{\bibfnamefont{Z.}~\bibnamefont{Wang}}, \bibnamefont{et~al.},
  \bibinfo{journal}{Phys. Rev. B} \textbf{\bibinfo{volume}{79}},
  \bibinfo{pages}{092507} (\bibinfo{year}{2009}).

\bibitem[{\citenamefont{Zhou et~al.}(2003)\citenamefont{Zhou, Lanzara,
  Bogdanov, Kellar, Shen, Yang, Ronning, Sasagawa, Kakeshita, Noda
  et~al.}}]{Zhou2003}
\bibinfo{author}{\bibfnamefont{X.~J.} \bibnamefont{Zhou}},
  \bibinfo{author}{\bibfnamefont{A.}~\bibnamefont{Lanzara}},
  \bibinfo{author}{\bibfnamefont{P.~V.} \bibnamefont{Bogdanov}},
  \bibinfo{author}{\bibfnamefont{S.~A.} \bibnamefont{Kellar}},
  \bibinfo{author}{\bibfnamefont{K.~M.} \bibnamefont{Shen}},
  \bibinfo{author}{\bibfnamefont{W.~L.} \bibnamefont{Yang}},
  \bibinfo{author}{\bibfnamefont{F.}~\bibnamefont{Ronning}},
  \bibinfo{author}{\bibfnamefont{T.}~\bibnamefont{Sasagawa}},
  \bibinfo{author}{\bibfnamefont{T.}~\bibnamefont{Kakeshita}},
  \bibinfo{author}{\bibfnamefont{T.}~\bibnamefont{Noda}}, \bibnamefont{et~al.},
  \bibinfo{journal}{Nature} \textbf{\bibinfo{volume}{423}},
  \bibinfo{pages}{398} (\bibinfo{year}{2003}).

\bibitem[{\citenamefont{Kondo et~al.}(2006)\citenamefont{Kondo, Takeuchi,
  Tsuda, and Shin}}]{Kondo2006}
\bibinfo{author}{\bibfnamefont{T.}~\bibnamefont{Kondo}},
  \bibinfo{author}{\bibfnamefont{T.}~\bibnamefont{Takeuchi}},
  \bibinfo{author}{\bibfnamefont{S.}~\bibnamefont{Tsuda}}, \bibnamefont{and}
  \bibinfo{author}{\bibfnamefont{S.}~\bibnamefont{Shin}},
  \bibinfo{journal}{Phys. Rev. B} \textbf{\bibinfo{volume}{74}},
  \bibinfo{pages}{224511} (\bibinfo{year}{2006}).

\bibitem[{\citenamefont{Eisaki et~al.}(2004)\citenamefont{Eisaki, Kaneko, Feng,
  Damascelli, Mang, Shen, Shen, and Greven}}]{Eisaki2004}
\bibinfo{author}{\bibfnamefont{H.}~\bibnamefont{Eisaki}},
  \bibinfo{author}{\bibfnamefont{N.}~\bibnamefont{Kaneko}},
  \bibinfo{author}{\bibfnamefont{D.~L.} \bibnamefont{Feng}},
  \bibinfo{author}{\bibfnamefont{A.}~\bibnamefont{Damascelli}},
  \bibinfo{author}{\bibfnamefont{P.~K.} \bibnamefont{Mang}},
  \bibinfo{author}{\bibfnamefont{K.~M.} \bibnamefont{Shen}},
  \bibinfo{author}{\bibfnamefont{Z.~X.} \bibnamefont{Shen}}, \bibnamefont{and}
  \bibinfo{author}{\bibfnamefont{M.}~\bibnamefont{Greven}},
  \bibinfo{journal}{Phys. Rev. B} \textbf{\bibinfo{volume}{69}},
  \bibinfo{pages}{064512} (\bibinfo{year}{2004}).

\bibitem[{\citenamefont{Fujita et~al.}(2005)\citenamefont{Fujita, Noda, Kojima,
  Eisaki, and Uchida}}]{Fujita2005}
\bibinfo{author}{\bibfnamefont{K.}~\bibnamefont{Fujita}},
  \bibinfo{author}{\bibfnamefont{T.}~\bibnamefont{Noda}},
  \bibinfo{author}{\bibfnamefont{K.~M.} \bibnamefont{Kojima}},
  \bibinfo{author}{\bibfnamefont{H.}~\bibnamefont{Eisaki}}, \bibnamefont{and}
  \bibinfo{author}{\bibfnamefont{S.}~\bibnamefont{Uchida}},
  \bibinfo{journal}{Phys. Rev. Lett.} \textbf{\bibinfo{volume}{95}},
  \bibinfo{pages}{097006} (\bibinfo{year}{2005}).

\bibitem[{\citenamefont{Romero et~al.}(1992)\citenamefont{Romero, Porter,
  Tanner, Forro, Mandrus, Mihaly, Carr, and Williams}}]{Romero1992}
\bibinfo{author}{\bibfnamefont{D.~B.} \bibnamefont{Romero}},
  \bibinfo{author}{\bibfnamefont{C.~D.} \bibnamefont{Porter}},
  \bibinfo{author}{\bibfnamefont{D.~B.} \bibnamefont{Tanner}},
  \bibinfo{author}{\bibfnamefont{L.}~\bibnamefont{Forro}},
  \bibinfo{author}{\bibfnamefont{D.}~\bibnamefont{Mandrus}},
  \bibinfo{author}{\bibfnamefont{L.}~\bibnamefont{Mihaly}},
  \bibinfo{author}{\bibfnamefont{G.~L.} \bibnamefont{Carr}}, \bibnamefont{and}
  \bibinfo{author}{\bibfnamefont{G.~P.} \bibnamefont{Williams}},
  \bibinfo{journal}{Phys. Rev. Lett.} \textbf{\bibinfo{volume}{68}},
  \bibinfo{pages}{1590} (\bibinfo{year}{1992}).

\bibitem[{\citenamefont{Tsvetkov et~al.}(1997)\citenamefont{Tsvetkov,
  Sch\"utzmann, Gorina, Kaljushnaia, and van~der Marel}}]{Tsvetkov1997}
\bibinfo{author}{\bibfnamefont{A.~A.} \bibnamefont{Tsvetkov}},
  \bibinfo{author}{\bibfnamefont{J.}~\bibnamefont{Sch\"utzmann}},
  \bibinfo{author}{\bibfnamefont{J.~I.} \bibnamefont{Gorina}},
  \bibinfo{author}{\bibfnamefont{G.~A.} \bibnamefont{Kaljushnaia}},
  \bibnamefont{and} \bibinfo{author}{\bibfnamefont{D.}~\bibnamefont{van~der
  Marel}}, \bibinfo{journal}{Phys. Rev. B} \textbf{\bibinfo{volume}{55}},
  \bibinfo{pages}{14152} (\bibinfo{year}{1997}).

\bibitem[{\citenamefont{Lupi et~al.}(2000)\citenamefont{Lupi, Calvani, Capizzi,
  and Roy}}]{Lupi2000}
\bibinfo{author}{\bibfnamefont{S.}~\bibnamefont{Lupi}},
  \bibinfo{author}{\bibfnamefont{P.}~\bibnamefont{Calvani}},
  \bibinfo{author}{\bibfnamefont{M.}~\bibnamefont{Capizzi}}, \bibnamefont{and}
  \bibinfo{author}{\bibfnamefont{P.}~\bibnamefont{Roy}},
  \bibinfo{journal}{Phys. Rev. B} \textbf{\bibinfo{volume}{62}},
  \bibinfo{pages}{12418} (\bibinfo{year}{2000}).

\bibitem[{\citenamefont{Lupi et~al.}(2009)\citenamefont{Lupi, Nicoletti, Limaj,
  Baldassarre, Ortolani, Ono, Ando, and Calvani}}]{Lupi2009}
\bibinfo{author}{\bibfnamefont{S.}~\bibnamefont{Lupi}},
  \bibinfo{author}{\bibfnamefont{D.}~\bibnamefont{Nicoletti}},
  \bibinfo{author}{\bibfnamefont{O.}~\bibnamefont{Limaj}},
  \bibinfo{author}{\bibfnamefont{L.}~\bibnamefont{Baldassarre}},
  \bibinfo{author}{\bibfnamefont{M.}~\bibnamefont{Ortolani}},
  \bibinfo{author}{\bibfnamefont{S.}~\bibnamefont{Ono}},
  \bibinfo{author}{\bibfnamefont{Y.}~\bibnamefont{Ando}}, \bibnamefont{and}
  \bibinfo{author}{\bibfnamefont{P.}~\bibnamefont{Calvani}},
  \bibinfo{journal}{Phys. Rev. Lett.} \textbf{\bibinfo{volume}{102}},
  \bibinfo{pages}{206409} (\bibinfo{year}{2009}).

\bibitem[{\citenamefont{Ando et~al.}(2000)\citenamefont{Ando, Hanaki, Ono,
  Murayama, Segawa, Miyamoto, and Komiya}}]{Ando2000}
\bibinfo{author}{\bibfnamefont{Y.}~\bibnamefont{Ando}},
  \bibinfo{author}{\bibfnamefont{Y.}~\bibnamefont{Hanaki}},
  \bibinfo{author}{\bibfnamefont{S.}~\bibnamefont{Ono}},
  \bibinfo{author}{\bibfnamefont{T.}~\bibnamefont{Murayama}},
  \bibinfo{author}{\bibfnamefont{K.}~\bibnamefont{Segawa}},
  \bibinfo{author}{\bibfnamefont{N.}~\bibnamefont{Miyamoto}}, \bibnamefont{and}
  \bibinfo{author}{\bibfnamefont{S.}~\bibnamefont{Komiya}},
  \bibinfo{journal}{Phys. Rev. B} \textbf{\bibinfo{volume}{61}},
  \bibinfo{pages}{014956(R)} (\bibinfo{year}{2000}).

\bibitem[{\citenamefont{Luo et~al.}(2008)\citenamefont{Luo, Cheng, Fang, and
  Wen}}]{Luo2008}
\bibinfo{author}{\bibfnamefont{H.}~\bibnamefont{Luo}},
  \bibinfo{author}{\bibfnamefont{P.}~\bibnamefont{Cheng}},
  \bibinfo{author}{\bibfnamefont{L.}~\bibnamefont{Fang}}, \bibnamefont{and}
  \bibinfo{author}{\bibfnamefont{H.~H.} \bibnamefont{Wen}},
  \bibinfo{journal}{Supercond. Sci. Technol.} \textbf{\bibinfo{volume}{11}},
  \bibinfo{pages}{125024} (\bibinfo{year}{2008}).

\bibitem[{\citenamefont{Homes et~al.}(1993)\citenamefont{Homes, Reedyk,
  Cradles, and Timusk}}]{Homes1993}
\bibinfo{author}{\bibfnamefont{C.~C.} \bibnamefont{Homes}},
  \bibinfo{author}{\bibfnamefont{M.}~\bibnamefont{Reedyk}},
  \bibinfo{author}{\bibfnamefont{D.~A.} \bibnamefont{Cradles}},
  \bibnamefont{and} \bibinfo{author}{\bibfnamefont{T.}~\bibnamefont{Timusk}},
  \bibinfo{journal}{Appl. Optics} \textbf{\bibinfo{volume}{32}},
  \bibinfo{pages}{2976} (\bibinfo{year}{1993}).

\bibitem[{\citenamefont{Allen}(1971)}]{Allen1971}
\bibinfo{author}{\bibfnamefont{P.~B.} \bibnamefont{Allen}},
  \bibinfo{journal}{Phys. Rev. B} \textbf{\bibinfo{volume}{3}},
  \bibinfo{pages}{305} (\bibinfo{year}{1971}).

\bibitem[{\citenamefont{Terasaki et~al.}(1990)\citenamefont{Terasaki, Tajima,
  Eisaki, Takagi, Uchinokura, and Uchida}}]{Terasaki1990}
\bibinfo{author}{\bibfnamefont{I.}~\bibnamefont{Terasaki}},
  \bibinfo{author}{\bibfnamefont{S.}~\bibnamefont{Tajima}},
  \bibinfo{author}{\bibfnamefont{H.}~\bibnamefont{Eisaki}},
  \bibinfo{author}{\bibfnamefont{H.}~\bibnamefont{Takagi}},
  \bibinfo{author}{\bibfnamefont{K.}~\bibnamefont{Uchinokura}},
  \bibnamefont{and} \bibinfo{author}{\bibfnamefont{S.}~\bibnamefont{Uchida}},
  \bibinfo{journal}{Phys. Rev. B} \textbf{\bibinfo{volume}{41}},
  \bibinfo{pages}{865} (\bibinfo{year}{1990}).

\bibitem[{\citenamefont{Puchkov et~al.}(1996)\citenamefont{Puchkov, Basov, and
  Timusk}}]{Puchkov1996}
\bibinfo{author}{\bibfnamefont{A.~V.} \bibnamefont{Puchkov}},
  \bibinfo{author}{\bibfnamefont{D.~N.} \bibnamefont{Basov}}, \bibnamefont{and}
  \bibinfo{author}{\bibfnamefont{T.}~\bibnamefont{Timusk}},
  \bibinfo{journal}{J. Phys.: Condens. Matter} \textbf{\bibinfo{volume}{8}},
  \bibinfo{pages}{10049} (\bibinfo{year}{1996}).

\bibitem[{\citenamefont{Yurgens et~al.}(2003)\citenamefont{Yurgens, Winkler,
  Claeson, Ono, and Ando}}]{Yurgens2003}
\bibinfo{author}{\bibfnamefont{A.}~\bibnamefont{Yurgens}},
  \bibinfo{author}{\bibfnamefont{D.}~\bibnamefont{Winkler}},
  \bibinfo{author}{\bibfnamefont{T.}~\bibnamefont{Claeson}},
  \bibinfo{author}{\bibfnamefont{S.}~\bibnamefont{Ono}}, \bibnamefont{and}
  \bibinfo{author}{\bibfnamefont{Y.}~\bibnamefont{Ando}},
  \bibinfo{journal}{Phys. Rev. Lett.} \textbf{\bibinfo{volume}{90}},
  \bibinfo{pages}{147005} (\bibinfo{year}{2003}).

\bibitem[{\citenamefont{Zheng et~al.}(2005)\citenamefont{Zheng, Kuhns, Reyes,
  Liang, and Lin}}]{Zheng2005}
\bibinfo{author}{\bibfnamefont{G.~Q.} \bibnamefont{Zheng}},
  \bibinfo{author}{\bibfnamefont{P.~L.} \bibnamefont{Kuhns}},
  \bibinfo{author}{\bibfnamefont{A.~P.} \bibnamefont{Reyes}},
  \bibinfo{author}{\bibfnamefont{B.}~\bibnamefont{Liang}}, \bibnamefont{and}
  \bibinfo{author}{\bibfnamefont{C.~T.} \bibnamefont{Lin}},
  \bibinfo{journal}{Phys. Rev. Lett.} \textbf{\bibinfo{volume}{94}},
  \bibinfo{pages}{047006} (\bibinfo{year}{2005}).

\bibitem[{\citenamefont{Ma et~al.}(2008)\citenamefont{Ma, Pan, Niestemski,
  Neupane, Xu, Richard, Nakayama, Sato, Takahashi, Luo et~al.}}]{Ma2008}
\bibinfo{author}{\bibfnamefont{J.~H.} \bibnamefont{Ma}},
  \bibinfo{author}{\bibfnamefont{Z.~H.} \bibnamefont{Pan}},
  \bibinfo{author}{\bibfnamefont{F.~C.} \bibnamefont{Niestemski}},
  \bibinfo{author}{\bibfnamefont{M.}~\bibnamefont{Neupane}},
  \bibinfo{author}{\bibfnamefont{Y.~M.} \bibnamefont{Xu}},
  \bibinfo{author}{\bibfnamefont{P.}~\bibnamefont{Richard}},
  \bibinfo{author}{\bibfnamefont{K.}~\bibnamefont{Nakayama}},
  \bibinfo{author}{\bibfnamefont{T.}~\bibnamefont{Sato}},
  \bibinfo{author}{\bibfnamefont{T.}~\bibnamefont{Takahashi}},
  \bibinfo{author}{\bibfnamefont{H.~Q.} \bibnamefont{Luo}},
  \bibnamefont{et~al.}, \bibinfo{journal}{Phys. Rev. Lett.}
  \textbf{\bibinfo{volume}{101}}, \bibinfo{pages}{207002}
  (\bibinfo{year}{2008}).

\bibitem[{\citenamefont{Kawasaki et~al.}(2010)\citenamefont{Kawasaki, Lin,
  Kuhns, Reyes, and Zheng}}]{Kawasaki2010}
\bibinfo{author}{\bibfnamefont{S.}~\bibnamefont{Kawasaki}},
  \bibinfo{author}{\bibfnamefont{C.}~\bibnamefont{Lin}},
  \bibinfo{author}{\bibfnamefont{P.~L.} \bibnamefont{Kuhns}},
  \bibinfo{author}{\bibfnamefont{A.~P.} \bibnamefont{Reyes}}, \bibnamefont{and}
  \bibinfo{author}{\bibfnamefont{G.~Q.} \bibnamefont{Zheng}},
  \bibinfo{journal}{Phys. Rev. Lett.} \textbf{\bibinfo{volume}{105}},
  \bibinfo{pages}{137002} (\bibinfo{year}{2010}).

\end{thebibliography}

\end{document}